\documentclass[aps,prb,twocolumn,amsmath,amssymb]{revtex4-1}

\usepackage{graphicx}
\usepackage{dcolumn}
\usepackage{bm}

\renewcommand{\vec}[1]{{\mathbf #1}}

\bibpunct{(}{)}{;}{s}{,}{,}

\begin{document}

\title{Interlayer Transport in Disordered Semiconductor Electron Bilayers}

\author{Youngseok Kim$^{1,2}$, Brian Dellabetta$^{1,2}$, and Matthew J. Gilbert$^{1,2}$}
\affiliation{$^1$Department of Electrical and Computer Engineering, University of Illinois, Urbana, Il, 61801}
\affiliation{$^2$Micro and Nanotechnology Laboratory, University of Illinois, Urbana, Il 61801}
\date{\today}

\begin{abstract}
We study the effects of disorder on the interlayer transport properties of disordered semiconductor bilayers outside of the quantum Hall regime by performing self-consistent quantum transport calculations. We find that the addition of material disorder to the system affects interlayer interactions leading to significant deviations in the interlayer transfer characteristics. In particular, we find that disorder decreases and broadens the tunneling peak, effectively reducing the interacting system to the non-interacting system, when the mean-free path for the electrons becomes shorter than the system length. Our results suggest that the experimental observation of exchange-enhanced interlayer transport in semiconductor bilayers requires materials with mean-free paths larger than the spatial extent of the system.
\end{abstract}


\maketitle

Semiconductor bilayer quantum wells have proven to be a system which exhibits an abundance of interesting physical phenomena. In this regard, semiconductor bilayers are mainly studied deep within the quantum Hall regime where strong interlayer interactions produce spectacular transport phenomena\cite{Spielman2000,Tiemann2009}. While semiconductor bilayers outside of the quantum Hall regime are not expected to undergo a dramatic phase transition, interlayer interactions have been predicted to produce interlayer transport phenomena similar to those characteristic of its quantum Hall regime though orders of magnitude smaller\cite{Swierkowski1997,cleanlimit}. While theoretically possible and potentially useful as a tunneling device\cite{Geppert2000}, to be experimentally observed, these exchange-enhanced interlayer transport signatures must be robust against material disorder which these semiconductor bilayer systems are well-known to contain. Therefore, understanding the observable deviations in interlayer transport caused by material disorder is required. 
\begin{figure}[b!]
  \centering
    \includegraphics[width=0.5\textwidth]{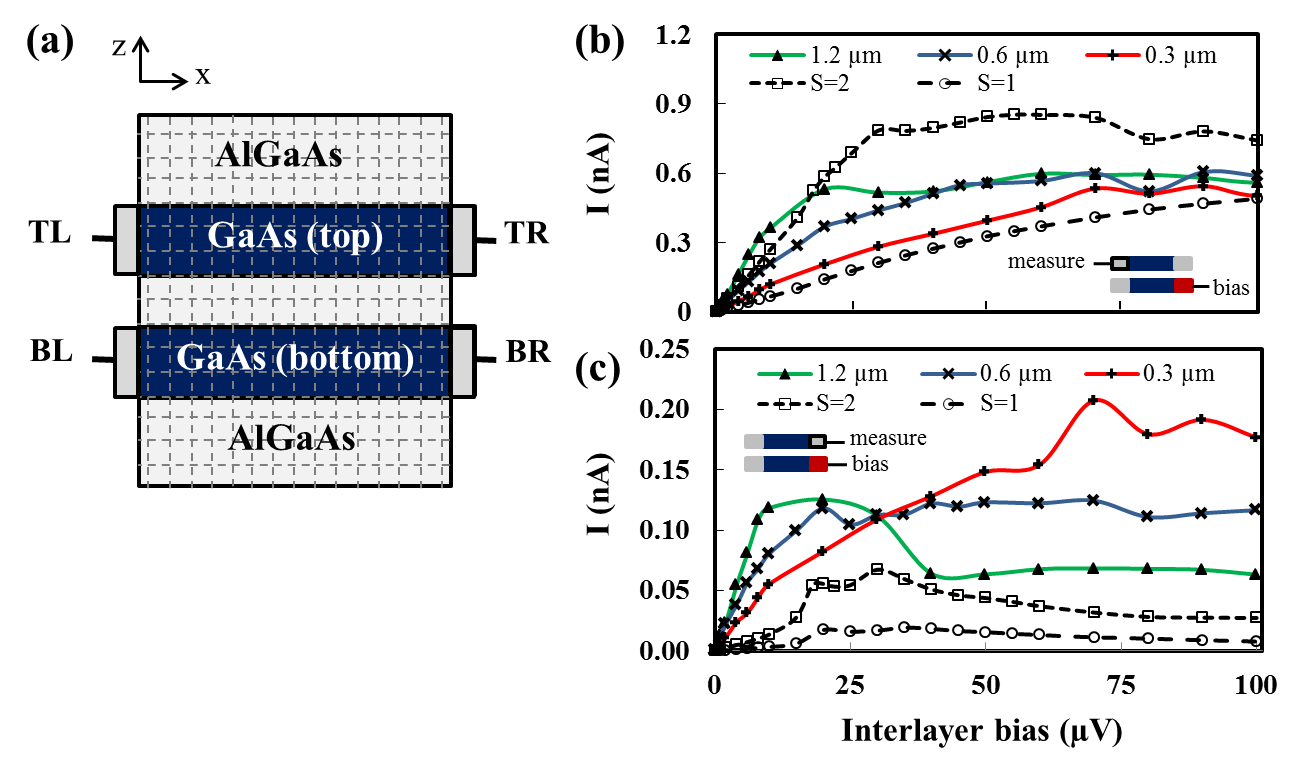}
  \caption{(Color online) (a): A schematic figure of the Al\(_{0.9}\)Ga\(_{0.1}\)As/GaAs system considered in this work. A plot of current as a function of interlayer bias for electrons injected into the bottom right contact and extracted from (b) the top left contact and (c) the top right contact. } \label{Fig:Device_IV}
\end{figure}
In Fig. \ref{Fig:Device_IV}(a), we plot a schematic of the system considered here. The system is comprised of two $16~nm$ deep GaAs quantum well separated by a $2~nm$ Al\(_{0.9}\)Ga\(_{0.1}\)As barrier. The gates on the top and bottom are isolated from their respective quantum well with $60~nm$ top and bottom Al\(_{0.9}\)Ga\(_{0.1}\)As barriers. Contacts inject and extract electrons from the system on the left and right ends of both the top and bottom layers. We use the tunneling bias configuration ($V_{TL}=V_{TR}=V_{BL}=0,~V_{BR}=V_{INT}$) in which electrons are injected into the bottom right contact and are extracted from the top left contact. The length and width of the system are each $1.2~\mu m$ and we assume that each 2DEG contains an electron concentration of $n_{2D}=2\times10^{10}~cm^{-2}$. The system temperature for all of our simulations is set to the zero temperature limit or $T_{sys}=0~K$. We define the single quantum well Hamiltonian as
\begin{equation}
\label{Hlayer}
H_{QW} = \sum\limits_{<i,j>} -\tau\mid i\rangle \langle j\mid + (4\tau + V_i)\mid i\rangle\langle i\mid,
\end{equation}
where lattice points $i$ and $j$ are nearest neighbors. $\tau=\hbar^2/2m^*a^2$ is the nearest neighbor hopping energy, where $m^*$ is the electron effective mass of GaAs and $a=20nm$ is the lattice constant of our simulation. $V_i=\phi(\vec{r}_i)$ is the on-site potential for GaAs from the normal tight-binding description and is calculated via a 3-dimensional Poisson solver. We may now generalize our 2DEG Hamiltonian to the double quantum well Hamiltonian by coupling the top and bottom quantum 2DEGs,
\begin{equation}
\label{SimpleHam}
{\cal{H}}_{sys} = \left[ \begin{array}{cc} H_{QW} & 0 \\ 0 & H_{QW} \end{array} \right] + \sum_{\mu = x,y,z} \hat{\mu} \cdot \vec{\Delta} \otimes \sigma_\mu ,
\end{equation}
with the interlayer interactions including both single particle tunneling and the mean-field many-body contribution, $\vec{\Delta}$, coupling the two quantum wells using a local density approximation\cite{Gilbert2010,cleanlimit}. In Eq. (\ref{SimpleHam}), $\mu$ represents a vector that isolates each of the Cartesian components of the pairing vector, and $\sigma_\mu$ represents the Pauli spin matrices in each of the three spatial directions. To complete our Hamiltonian definition, 
we use $t=\Delta_{sas}/2=2\mu eV$ as the single particle tunneling amplitude. We model the exchange interaction using an enhancement factor\cite{Swierkowski1997} $S$, which effectively gives the quasi-particle tunneling amplitude as $t_{eff}=S t$. Therefore, $\vec{\Delta}$ in Eq. (\ref{SimpleHam}) is a function of $t$ and $S$\cite{Gilbert2010,cleanlimit}. The predicted value for $S$ is $\sim2$ and we use $S=2$ for the remainder of this work when discussing interacting bilayer systems.

To understand the effects of material disorder in the GaAs/AlGaAs system\cite{Ando1991}, the impurities are assumed to be highly screened in the 2DEG. As a result, the impurity potential can be described as a $\delta$-function. Within a lattice model, random disorder can be inserted into the system Hamiltonian by adding a uniform distribution having an energy window of $W$ to the on-site energy. The $W$ in two dimensional lattice model containing high concentrations of $\delta$-function impurities satisfies the relation\cite{Ando1991}
\begin{equation}
\label{Eq:WLambda}
\frac{W}{E_F}=\left(\frac{6\lambda_F^3}{\pi^3a^2\Lambda}\right)^{1/2},
\end{equation}
where $E_F$, $\lambda_F$ are Fermi energy and Fermi wavelength, respectively, and $a$ is a lattice constant. Consequently, the impurity level is now characterized by a mean free path (MFP), $\Lambda$. We do not find any signficant contribution to the interlayer transport when the selected MFP is approximately as large as the system length. Thus, we only focus on the MFP comparable to or smaller than the device size in this paper. The calculation strategy follows a standard self-consistent field procedure. Given the density matrix of the 2D system, we can evaluate the mean-field quasiparticle Hamiltonian with the non-equilibrium Green's function (NEGF) formalism coupled with a 3D Poisson equation. We calculate all transport properties via Landauer formula and all resultant values are averaged over at least four different disorder configurations per MFP.

In Fig. \ref{Fig:Device_IV}(b) and \ref{Fig:Device_IV}(c), we analyze the $I-V$ characteristics of the disordered system sweeping the interlayer bias up to $100\mu V$. Here the critical current, or amount of interlayer current the system can sustain by simply altering the interlayer phase, from the simulation is $I_c^{clean}=0.89nA$, which is smaller than the predicted critical current\cite{cleanlimit} of $I_c^{anal}=1.44nA$. This discrepancy can be understood by considering the charge imbalance and efficient interlayer charge screening by the electrons, especially at high bias window where the quantum wells are off-resonance. We now introduce impurities into the system corresponding to three different MFPs of $\Lambda=1.2,~0.6$, and $0.3\mu m$. The range of MFPs we consider here adds a significant perturbation to the on-site energy of $10\sim20\%$ of the self-consistent Hartree potential. In Fig. \ref{Fig:Device_IV}(b), we plot the current flowing from the bottom right contact (BR) to the top left contact (TL) and immediately see the effects of even small amounts of disorder. By comparing the interacting and non-interacting systems, we notice that the interacting system has a uniform current enhancement over the non-interacting case over all interlayer voltages. However, when disorder corresponding to a MFP commensurate with the system size is introduced, we see that the interlayer current is increased at low $V_{INT}$. This is a consequence of local regions of stronger interlayer interactions set up by disorder-induced localization. As $V_{INT}$ is increased, the interlayer enhancement quickly disappears and leaves only a small exchange enhancement over the non-interacting case. As the MFP is further decreased, we see that the trend of decreased interlayer exchange enhancement continues as the perturbations to the carriers via the disorder potential is now large enough to energetically separate the two layers leaving only weak interlayer interactions.
A separate way of visualizing the effects of disorder is via Fig. \ref{Fig:Device_IV}(c) where we plot the the current injected from the bottom right (BR) contact and extracted from the top right contact (TR). Here, we would expect the interlayer current to be close to zero as we see for the $S=1$ and $S=2$ cases without disorder. However, when disorder localizes the electron states in the bottom layer, interlayer tunneling is locally enhanced leading to increased interlayer transmission near the injection contact. Beyond locally enhanced transmission, the presence of disorder in the top layer perturbs the electron states leading to backscattering of injected states. The overall effect is an increasing current from BR to TR as disorder increases.

\begin{figure}[t!]
  \centering
    \includegraphics[width=0.5\textwidth]{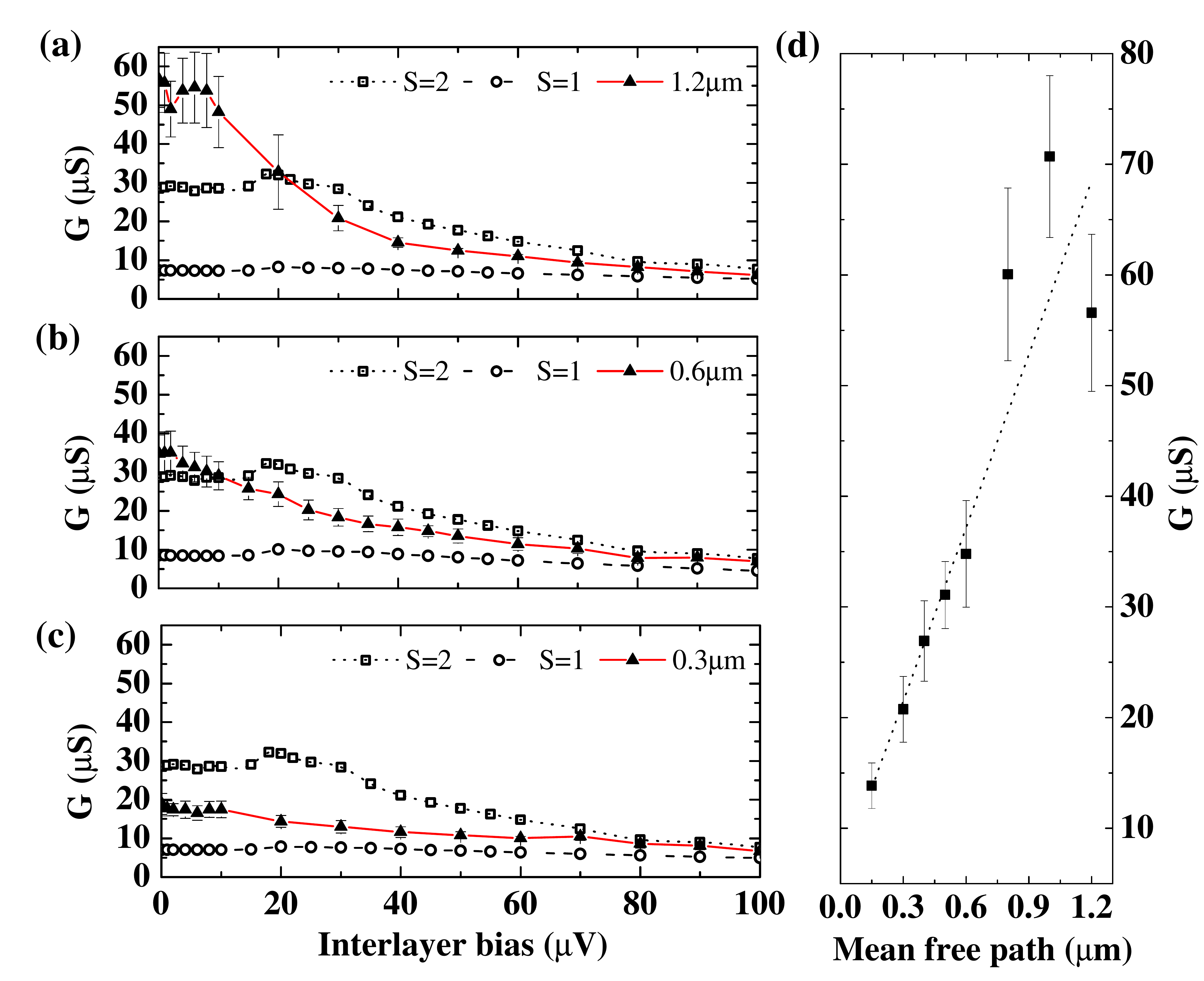}
  \caption{(Color online) Plots of a interlayer conductance, $G$, as a function of interlayer bias with mean free path of (a) $1.2\mu m$, (b) $0.6\mu m$, and (c) $0.3\mu m$. (d) Plot of max height of the interlayer conductance as a function of mean free path at $V_{INT}=0.1\mu V$. } \label{Fig:dIdV}
\end{figure}
\begin{figure}[t!]
  \centering
    \includegraphics[width=0.5\textwidth]{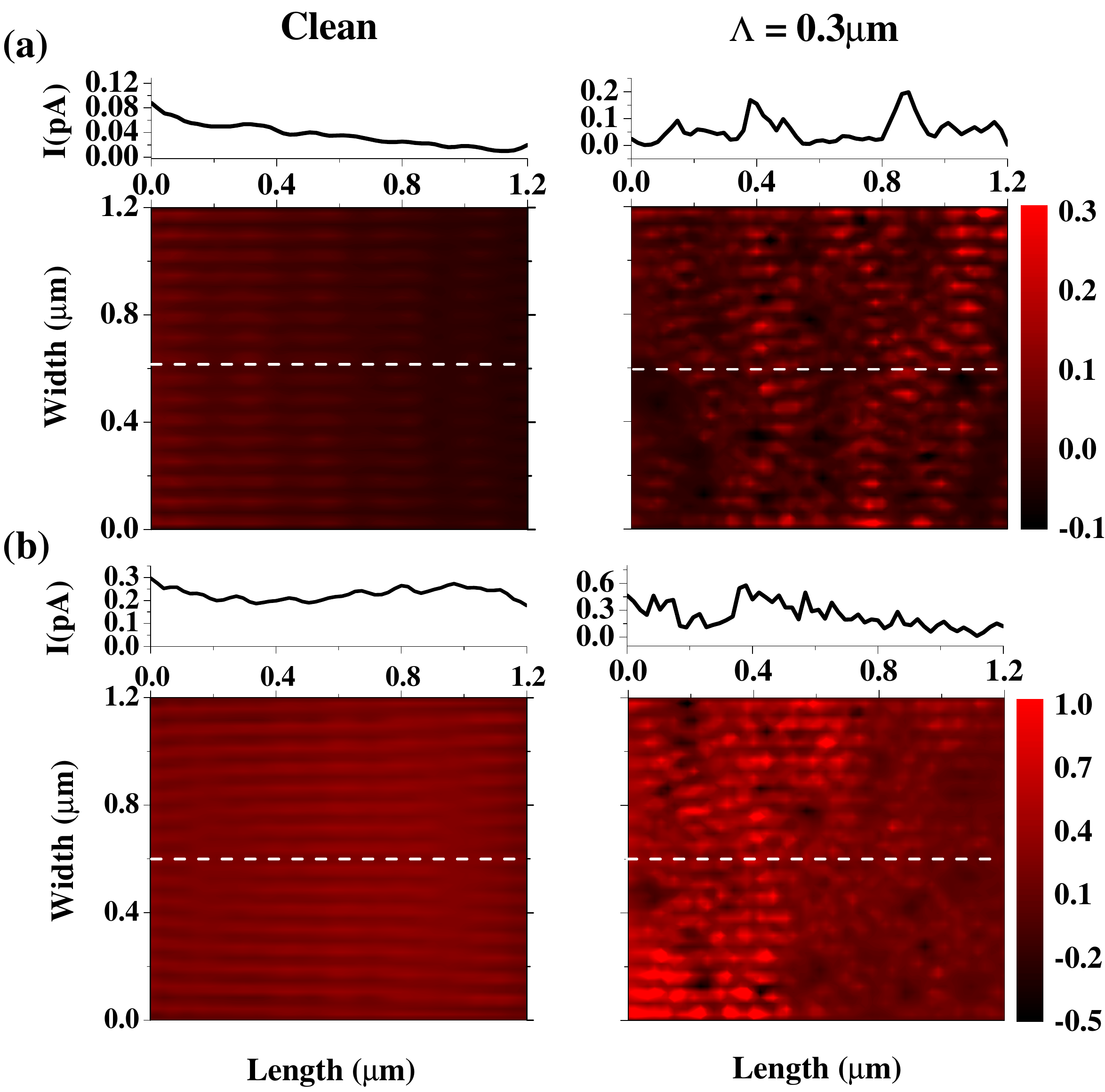}
  \caption{(Color online) A plot of spatially resolved current as a function of device width and length at (a) $V_{INT}=10\mu V$ and (b) $V_{INT}=80~\mu V$ in units of pA for an assumed enhancement factor of $S=2$. The current profile of the clean system (disordered system of $\Lambda=0.3\mu m$) is presented at left (right) column. Above each two dimensional plot, there is a current profile taken in the middle of the system indicated as a white dashed line. } \label{Fig:IV}
\end{figure}
A crucial component to understanding the interlayer dynamics in the presence of materials disorder is the critical current, or the interlayer current at which point the system switches from coherent exchange-enhanced electron dynamics to incoherent single-particle dynamics. As we have not passed a phase boundary, we do not expect large differences in interlayer currents between the two transport regimes. Examination of the differential conductance is well understood to provide an excellent probe of the drop in interlayer conductivity associated with reaching critical current\cite{Tiemann2009}. We calculate a critical current for each disorder configuration and MFP and find that $I_c^{1.2\mu m}=0.67\pm 0.02nA$, $I_c^{0.6\mu m}=0.72\pm 0.09nA$, and $I_c^{0.3\mu m}=0.74\pm 0.07nA$ which is smaller than the disorder-free case ($I_c^{clean}=0.89nA$). However, we find the critical current saturates with increasing disorder but this is simply because the exchange enhancement has been lost and the system responds as if it is non-interacting for MFPs shorter than the system dimensions. 

Figures \ref{Fig:dIdV}(a) - \ref{Fig:dIdV}(c) depict tunneling conductance which is defined as $G=\langle I_{TL}+I_{TR} \rangle/V_{INT}$, and show that the width of the tunneling conductance peak is broadened with the increasing material disorder. With a given MFP, the broadening (Lorentzian half-width, $\Gamma$) is calculated as\cite{Turner1996} $\Gamma=\frac{\hbar}{\tau}$, where $\tau$ is quantum lifetime of the electrons. The predicted broadening ($\Gamma/2$) for $\Lambda=1.2,~ 0.6,~ 0.3\mu m$ are around $20, ~41, ~82\mu eV$, respectively. Our results show energy broadening of $22.1\pm 2.0,~ 36.7\pm 5.8,~ 76.0\pm 13.7\mu eV$, respectively, in a good agreement with the predicted values. In Fig. \ref{Fig:dIdV}(d), we plot the maximum value of the interlayer conductance as a function of MFP. We observe a linear dependence in the differential interlayer conductance\cite{Zheng1993}, especially at the MFP shorter than $\Lambda=0.6\mu m$. The broadening in tunneling peak and linear dependence of interlayer conductance in MFP indicate that the interlayer transport is no longer controlled by the intrinsic nature of the system and instead is controlled by the material disorder when the mean free path is shorter than device length. By reducing the MFP to $\Lambda=0.3\mu m$, we find another overall reduction of the interlayer conductance peak value with additional scattering enhanced broadening of the interlayer conductance peak. 

As important as it is to understand the variations in the magnitude of the interlayer currents, we must also understand how disorder shifts the interlayer current density. In Fig. \ref{Fig:IV}, we plot the interlayer current density below and above critical current for a clean system and one with a MFP of $\Lambda=0.3~\mu m$. In the linear response regime, the tunneling current\cite{cleanlimit} is proportional to $v_f^{-2}$, where $v_f$ is a quasi-particle Fermi velocity in transport direction. The quasi-particle subband with the highest transverse velocity has the lowest transport directional velocity and, consequently, makes the largest contribution to the interlayer transport. As a result, the quasi-particle wavefunction of the highest quantized level contributes most significantly to the tunneling current, and is manifest as a transverse directional oscillation in the current profile in Fig. \ref{Fig:IV} due to the small degree of confinement in our $1.2\mu m$ wide structure. In Fig. \ref{Fig:IV}(a), we see that in the clean case, at small interlayer bias, the interlayer current density is peaked near the contacts but flows randomly across the system in the presence of strong disorder. This is clearly not the case in Fig. \ref{Fig:IV}(b), where we see that above critical current the clean system exhibits an even spatial distribution of the interlayer current but that the disordered system has been broken up into regions of strong and weak interlayer current flow controlled by the disorder pattern. This gives further evidence of how disorder with a corresponding MFP smaller than the system dimensions controls the interlayer transport properties. 

We have shown that while exchange-enhanced interlayer transport properties in semiconductor bilayers may exist outside of the quantum Hall regime they are very sensitive to the strength and location of material disorder. We find that as soon as the MFP of the electrons becomes commensurate with the spatial system dimensions that the disorder controls the interlayer transport properties and reduces the interacting electron system to a non-interacting one. These findings demonstrate that to observe these states experimentally one requires systems with MFPs greater than the system spatial dimensions.

\begin{acknowledgements}
This work is supported by the Army Research Office (ARO) under Contract No. W911NF-09-1-0347. We acknowledge support for the Center for Scientific Computing from the CNSI, MRL: an NSF MRSEC (DMR-1121053) and NSF CNS-0960316 and Hewlett-Packard.  Y. Kim is supported by Fulbright Science and Technology Award.
\end{acknowledgements}

\bibliographystyle{apsrev}
\bibliography{manuscript_letter}		

\end{document}